\documentclass{tlp}

\usepackage{amsmath,amssymb}
\usepackage{times,helvet,courier}
\usepackage{graphicx}
\usepackage[boxed,vlined]{algorithm2e}
\usepackage{dsfont}
\usepackage{subfig}
\usepackage{tikz}
\usetikzlibrary{shapes,arrows}
\usepackage{listings}

\usepackage{url} 

\title{ASP modulo CSP: The \textit{clingcon} system}
\author[Max Ostrowski \and Torsten Schaub]{%
  Max Ostrowski
  and
  Torsten Schaub 
  \\
  Institut f\"ur Informatik, Universit\"at Potsdam}

\begin{document}

\newcommand{\claspD}{\textit{claspD}}
\newcommand{\claspar}{\textit{claspar}}
\newcommand{\claspfolio}{\textit{claspfolio}}
\newcommand{\claspre}{\textit{claspre}}
\newcommand{\clasp}{\textit{clasp}}
\newcommand{\clingcon}{\textit{clingcon}}
\newcommand{\clingo}{\textit{clingo}}
\newcommand{\coala}{\textit{coala}}
\newcommand{\gringo}{\textit{gringo}}
\newcommand{\iclingo}{\textit{iclingo}}
\newcommand{\plasp}{\textit{plasp}}

\newcommand{\dlv}{\textit{dlv}}
\newcommand{\lparse}{\textit{lparse}}
\newcommand{\smodels}{\textit{smodels}}

\newcommand{\lua}{\textit{lua}}
\newcommand{\gecode}{\textit{gecode}}

\newcommand{\Tsign}{\ensuremath{\mathbf{T}}}
\newcommand{\Fsign}{\ensuremath{\mathbf{F}}}
\newcommand{\Tlit}[1]{\Tsign \lit{#1}}
\newcommand{\Flit}[1]{\Fsign \lit{#1}} 
\newcommand{\inv}[1]{\ensuremath{\overline{#1}}}
\newcommand{\catoms}{\ensuremath{\mathcal{C}}}
\newcommand{\Tass}[1]{\ensuremath{#1^{\Tsign}}}
\newcommand{\Fass}[1]{\ensuremath{#1^{\Fsign}}}

\newcommand{\Tcon}[1]{\ensuremath{\Tsign}{\con{#1}}}
\newcommand{\Fcon}[1]{\ensuremath{\Fsign}{\con{#1}}}

\newcommand{\Tmylit}[1]{\Tsign \mylit{#1}}
\newcommand{\Fmylit}[1]{\Fsign \mylit{#1}} 

\newcommand{\con}[1]{\texttt{#1}}
\newcommand{\mylit}[1]{\texttt{#1}}
\newcommand{\lit}[1]{\ensuremath{#1}}
\newcommand{\cvar}[1]{\texttt{#1}}

\newcommand{\naf}[1]{\ensuremath{\mathit{not}~{#1}}}
\newcommand{\atom}[1]{\ensuremath{\mathit{atoms}(#1)}}

\newcommand{\project}[2]{\ensuremath{{#1}|_{#2}}}

\newcommand{\myparagraph}[1]{\par\emph{#1}.}

\RestyleAlgo{algoruled}
\DontPrintSemicolon
\LinesNumbered
\SetAlgoVlined
\SetFuncSty{textsc}

\SetKwInOut{Input}{Input}
\SetKwInOut{Output}{Output}
\SetKw{Return}{return}


\newlength\listingnumberwidth
\setlength\listingnumberwidth{17.0pt}
\lstset{numbers=left,numberblanklines=false,basicstyle=\ttfamily\scriptsize,xleftmargin=\listingnumberwidth}
\lstset{escapeinside={(*@}{@*)}}
\lstset{keywords=[1]{\$, \$domain, \$maximize, \$minimize, \$count, \$distinct},keywordstyle=[1]\textbf%
,keywords=[2]{work(A), work(B), work(adam), work(lea), work(smith), work(john), fulltime},keywordstyle=[2]\usefont{OT1}{cmtt}{m}{n}%
,alsoletter={\$,\(,\)}
,alsoother={\@,\,,\$,\(}}
\lstset{morecomment=[l]\%,commentstyle=\textit}


\newcommand{\vertgap}{\smallskip}
\newcommand{\moregap}{\hspace{-3.5mm}}
\newcommand{\backgap}{\hspace{-3mm}}
\newcommand{\lackgap}{\hspace{-2mm}}
\newcommand{\code}[1]{\lstinline[escapeinside={(*@}{@*)}]!#1!}
\newcommand{\codeeq}[1]{\lstinline[basicstyle=\tt\small]{\ \ #1}}
\newcommand{\breakeq}{\\[-1.5pt]}


\maketitle

\begin{abstract}
We present the hybrid ASP solver \clingcon,
combining the simple modeling language and the high performance Boolean solving
capacities of Answer Set Programming (ASP) with techniques for using non-Boolean
constraints from the area of Constraint Programming (CP).
The new \clingcon\ system features an extended syntax supporting global
constraints and optimize statements for constraint variables.
The major technical innovation improves the interaction between ASP and CP
solver through elaborated learning techniques based on
\emph{irreducible inconsistent sets}.
A broad empirical evaluation shows that these
techniques yield a performance improvement of an order of magnitude.
To appear in Theory and Practice of Logic Programming.
\end{abstract}


\section{Introduction}\label{sec:introduction}

\textit{clingcon} is a hybrid solver for Answer Set Programming (ASP;~\cite{baral02a}),
combining the simple modeling language and the high performance Boolean solving
capacities of ASP with techniques for using non-Boolean constraints from the
area of Constraint Programming (CP).
Although \clingcon's solving components follow the approach of modern
Satisfiability Modulo Theories (SMT;~\cite[Chapter~26]{SATHandbook})
solvers when combining the ASP solver \textit{clasp} with the CP solver
\textit{gecode}~\cite{gecode},
\clingcon\ furthermore adheres to the tradition of ASP in supporting a
corresponding modeling language by appeal to the ASP grounder \textit{gringo}.
Although in the current implementation the theory solver is instantiated with
the CP solver \textit{gecode},
the principal design of \clingcon\ along with the corresponding interfaces are
conceived in a generic way, aiming at arbitrary theory solvers.

The underlying formal framework, 
defining syntax and semantics of constraint logic programs, 
and the principal algorithms, 
were presented in~\cite{geossc09a}.
This initial \textit{clingcon} system~0.1.0 was based on \textit{clingo}~2.0.2
and \textit{gecode}~2.2.0.
Unlike this,
the new version of \textit{clingcon} is based on \textit{clingo}~3.0.4
and \textit{gecode}~3.7.1.
Apart from major refactoring,
it features an extended syntax supporting global constraints and optimize
statements for constraint variables.
Also, it allows for more fine-grained configurations of constraint-based
lookahead, optimization, and propagation delays.
However, the major technical innovation improves the interaction between ASP and CP
solver through elaborated learning techniques.
We introduce filtering methods for conflicts and reasons based on
\emph{irreducible inconsistent sets}.
A broad empirical evaluation shows that these
techniques yield a performance improvement of an order of magnitude.


\section{The \clingcon\ approach}\label{sec:background}

The input language of \clingcon\ extends the one of \emph{gringo}~\cite{potasscoManual} 
by CP-specific operators marked with a preceding \texttt{\$} symbol (cf. Section~\ref{sec:language}).
After grounding, a propositional program is then composed of regular and
constraint atoms, denoted by $\mathcal{A}$ and $\mathcal{C}$, respectively.
The set of constraint atoms induces an ordinary constraint satisfaction problem (CSP)
$(V,D,C)$,
where
$V$ is a set of variables
with common domain~$D$,
and $C$ is a set of constraints.
This CSP is to be addressed by the corresponding CP solver.
As detailed in~\cite{geossc09a},
the semantics of such constraint logic programs is defined by appeal to a
two-step reduction.
For this purpose,
we consider
a regular Boolean assignment over $\mathcal{A}\cup\mathcal{C}$ 
(in other words, an interpretation) and 
an assignment of $V$ to $D$ 
(for interpreting the variables $V$ in the underlying CSP).
In the first step,
the constraint logic program is reduced to a regular logic program by
evaluating its constraint atoms.
To this end, the constraints in $C$ associated with the program's constraint
atoms $\mathcal{C}$ are evaluated w.r.t. the assignment of $V$ to $D$.
In the second step, the common Gelfond-Lifschitz reduct~\cite{gellif91a} is performed
to determine whether the Boolean assignment is an answer set of the obtained
regular logic program.
If this is the case, the two assignments constitute a (hybrid) constraint answer
set of the original constraint logic program.

In what follows,
we rely upon the following terminology.
We use signed literals of form \Tsign{a} and \Fsign{a} 
to express that an atom $a$ is assigned \Tsign\ or \Fsign,
respectively.
That is, \Tsign{a} and \Fsign{a} stand for the Boolean assignments
$a\mapsto\Tsign$ and $a\mapsto\Fsign$, respectively.
We denote the complement of such a literal~$\ell$ by $\overline{\ell}$.
That is,
$\overline{\Tsign{a}}=\Fsign{a}$ and $\overline{\Fsign{a}}=\Tsign{a}$.
We represent a Boolean assignment simply by a set of signed literals.
Sometimes we restrict such an assignment $A$ to its regular or constraint atoms
by writing $\project{A}{\mathcal{A}}$ or $\project{A}{\mathcal{C}}$,
respectively.
For instance,
given the regular atom `\mylit{person(adam)}' and the constraint atom `\con{work(adam)~\$>~4}',
we may form the Boolean assignment
\(
\{\Tmylit{person(adam)},\Fcon{work(adam)~\$>~4}\}
\).

We identify constraint atoms in $\mathcal{C}$ with constraints in $(V,D,C)$ via
a function
\(
\gamma: \mathcal{C}\to C
\).
Provided that each constraint $c\in C$ has a complement $\overline{c}\in C$,
like $\overline{\text{`}x=y\text{'}}={\text{`}x\neq y\text{'}}$ or $\overline{\text{`}x<y\text{'}}={\text{`}x\geq y\text{'}}$ and vice versa,
we extend $\gamma$ to signed constraint atoms over $\mathcal{C}$:
\[
\gamma(\ell)
=
\left\{
  \begin{array}{ll}
              c  & \text{ if } \ell=\Tsign{c}
    \\
    \overline{c} & \text{ if } \ell=\Fsign{c}
  \end{array}
\right.
\]
For instance, we get
\(
\gamma(\Fcon{work(adam)~\$>~4})
=
work(adam)\leq 4
\),
where $\cvar{work(adam)}\in V$ is a constraint variable and
$(work(adam)\leq 4)\in C$ is a constraint.
An assignment satisfying the last constraint is $\{work(adam)\mapsto 3\}$.

Following~\cite{gekanesc07a},
we represent Boolean constraints issuing from a logic program under ASP semantics
in terms of \emph{nogoods}~\cite{dechter03}.
This allows us to view inferences in ASP as unit propagation on
nogoods.
A \emph{nogood} is a set
$\{\sigma_1,\dots,\sigma_m\}$ of signed literals,
expressing that any assignment containing $\sigma_1,\dots,\sigma_m$ is unintended.
Accordingly,
a total assignment $A$ is a \emph{solution} for a set~$\Delta$ of nogoods
if $\delta\not\subseteq A$ for all $\delta\in\Delta$.
Whenever $\delta\subseteq A$, the nogood $\delta$ is said to be
\emph{conflicting} with $A$.
For instance,
given atoms $a,b$,
the total assignment $\{\Tsign{a},\Fsign{b}\}$
is a solution for the set of nogoods containing
$\{\Tsign{a},\Tsign{b}\}$ and $\{\Fsign{a},\Fsign{b}\}$.
Likewise, $\{\Fsign{a},\Tsign{b}\}$ is another solution.
Importantly, nogoods provide us with reasons explaining why entries
must (not) belong to a solution,
and lookback techniques can be used to analyze and recombine 
inherent reasons for conflicts.
We refer to~\cite{gekanesc07a} on how logic
programs are translated into nogoods within ASP.


\section{The \clingcon\ Architecture}\label{sec:architecture}

Although \clingcon's solving components follow the approach of modern SMT
solvers when combining the ASP solver \textit{clasp} with the CP solver
\textit{gecode},
\clingcon\ furthermore adheres to the tradition of ASP in supporting a
corresponding modeling language based on the ASP grounder \textit{gringo}.
The resulting tripartite  architecture of \clingcon\ is depicted in
Figure~\ref{fig:architecture}.
\begin{figure}
\tikzstyle{system}=[draw, text width=5em, 
    text centered , minimum height=5.5em, minimum width=2.5cm]

\tikzstyle{extension}=[draw, text width=5em, 
    text centered, minimum height=1.5em]
		\small
\pgfdeclarelayer{background}
\pgfdeclarelayer{foreground}
\pgfsetlayers{background,main,foreground}

\begin{tikzpicture}
\begin{pgfonlayer}{background}
\node (gringo) [system]  {};
\path (gringo.east)+(3.2,0) node (clasp) [system] {};
\path [draw, ->] (gringo.east) -- node [above] {} 
         (clasp.west) ;
\end{pgfonlayer}
\path (gringo.south)+(0,0.6) node (languageext) [extension] {Theory\\Language};
\path (gringo)+(0,0.6) node (gringotext) [] {\textit{gringo}};
\path (clasp)+(0,0.6) node (clasptext) [] {\textit{clasp}};
\path (clasp.south)+(0,0.6) node (theorypropagator) [extension] {Theory\\Propagator};

\path (theorypropagator.east)+(3.2,0) node (theorysolver) [extension] {Theory\\Solver};

\path [draw, <->] (theorypropagator.east) -- node [above] {} 
         (theorysolver.west) ;
\path [draw, ->] (languageext.east) -- node [above] {} 
         (theorypropagator.west) ;
\end{tikzpicture}
\caption{Architecture of \clingcon}
\label{fig:architecture}
\end{figure}
%
Although in the current implementation the theory solver is instantiated with
the CP solver \textit{gecode},
the principal design of \clingcon\ along with the corresponding interfaces are
conceived in a generic way, aiming at arbitrary theory solvers.

Following the workflow in Figure~\ref{fig:architecture},
the first extension concerns the input language of \textit{gringo}
with theory-specific language constructs.
Just as with regular atoms, the grounding capabilities of \textit{gringo} can be
used for dealing with constraint atoms containing first-order variables.
As regards the current \textit{clingcon} system,
the language extensions allow for expressing constraints over integer variables.
As we detail in Section~\ref{sec:language}, this involves arithmetic constraints
as well as global constraints and optimization statements.
These constraints are treated as atoms and passed to \clasp\ via the
standard \gringo-\clasp\ interface,
also used in \clingo, the monolithic combination of \gringo\ and \clasp.
Information about these constraints is furthermore directly shared with the
theory propagator and in turn the theory solver, viz.\ \textit{gecode}.
In the new version of \textit{clingcon},
the theory propagator is implemented as a post propagator,
as furnished by \textit{clasp}.\footnote{
  Post propagators provide an abstraction easing \textit{clasp}'s extensibility with more
  elaborate propagation mechanisms.
  To this end,
  \textit{clasp} maintains a list of post propagators that are consecutively
  processed after unit propagation.
  Also,
  lookahead and unfounded-set checking are implemented in \textit{clasp} as post
  propagators.}
Theory propagation is done in the theory solver until a fixpoint is
reached.
In doing so,
decided constraint atoms are transferred to the theory solver,
and conversely constraints whose truth values are determined by the theory solver
are sent back to \textit{clasp} using a corresponding nogood.
Note that theory propagation is not only invoked when propagating partial
assignments but also whenever a total Boolean assignment is found.
Whenever the theory solver detects a conflict,
the theory propagator is in charge of conflict analysis.
Apart from reverting the state of the theory solver upon backjumping,
this involves the crucial task of determining a conflict nogood 
(which is usually not provided by theory solvers, as in the case of \textit{gecode}).
This is elaborated upon in Section~\ref{sec:minimization}.
Similarly,
the theory propagator is in charge of enumerating constraint variable
assignments, whenever needed.
Finally,
we note that the theory propagator is informed whenever constraint
atoms are decided.
This allows for updating watches for constraint literals,
even before propagation is launched.
This is implemented by \textit{clasp}'s call-back routines,
another new feature of \textit{clasp} supporting theory solving.


\section{The \clingcon\ Language}\label{sec:language}

We explain the syntax of our constraint logic programs
via the example in Listing~\ref{listing:example1}.
\begin{lstlisting}[float=t%
,frame=lines%
,framesep=0pt%
,caption={Example of a constraint logic program}%
%          (\accessCumulative)}% : \lstinline{oclingo --iinit="1-offset"\ \accessCumulative}
,label=listing:example1%
%,linerange={14-16,22-23,25-26,28-29,35-36,39-39,42-45,48-48,54-55,58-58}%,63-66}%
,linerange={0-15}%,63-66}%
]
$domain(*@@*)(0..10).(*@\label{line:domain}@*)
person(adam;smith;lea;john).
1{team(A,B) : person(B) : B != A}1 :- person(A), A == adam.(*@\label{line:choose}@*)
{friday}.

work(A) $+ work(B) $> 6 :- team(A,B).(*@\label{line:gt6}@*)
work(B) $- work(adam) $== 1 :- friday, team(adam,B).(*@\label{line:daughter}@*)
:- team(adam,lea), not work(lea) $== work(adam).(*@\label{line:couple}@*)
work(B) $== 0 :- person(B), not team(adam,B), B != adam.(*@\label{line:prevent}@*)

$count[work(A) $== 8 : person(A)] $== fulltime.(*@\label{line:global}@*)

$maximize{work(A) : person(A)}.(*@\label{line:optimize}@*)
\end{lstlisting}
Suppose Adam wants to do house renovation with
the help of three friends.
We encode the problem as follows.
In Line~\ref{line:domain} we restrict all constraint variables
to the domain $[0,10]$ as nobody wants to work more than 10 hours a day.
In Line~\ref{line:choose} we choose teams.
They agreed that each team has to work more than six hours a day (Line~\ref{line:gt6}).
Within this line we show the syntax of linear constraints.
They can be used in the head or body of a rule\footnote{Constraint atoms in the head are shifted to the negative body.}.
We use the \texttt{\$} sign in front of every relation and function symbol
referring to the underlying CSP.
In this new {\clingcon} version this also applies to arithmetic operators
to better separate them from {\gringo} operators.
Many standard arithmetical operators are supported, like plus($+$),  times($*$) and absolute($abs$).
We use the grounding capabilities of {\gringo} to create the constraint variables.
Grounding Line~\ref{line:gt6} yields:
\begin{lstlisting}[numbers=none]
work(adam) $+ work(smith) $> 6 :- team(adam,smith).
work(adam) $+ work(lea)   $> 6 :- team(adam,lea).
work(adam) $+ work(john)  $> 6 :- team(adam,john).
\end{lstlisting}
We created three ground rules containing three different constraints,
using four different constraint variables.
Note that the constraint variables have not been defined beforehand.
All variables occurring in a constraint are automatically constraint variables.
On Fridays, Adam has to pick up his daughter from sports and
therefore works one hour less than his partner (Line~\ref{line:daughter}).
Furthermore Lea and Adam are a couple and decided to have an equal work load
if they are in the same team (Line~\ref{line:couple}).
Finally,
Line~\ref{line:prevent} prevents persons from working
if they are not teammates.

With this little constraint logic program,
we want to show how for example quantities can be easily expressed.
Constraints and constraint variables fit naturally into the logic program.
Not representing quantities explicitly with propositional variables
eases the modelling of problems and also decreases the size of the ground
logic program.

\myparagraph{Global Constraints}
In Line~\ref{line:global} we use a global constraint.
This is a new feature of {\clingcon}.
Global constraints capture relations
between a non-fixed number of variables.
As with aggregates in  {\gringo},
we can use conditional literals~\cite{lparseManual}
to represent these sets of variables.
In the example we can see a \emph{count} constraint.
Grounding this yields:
\begin{lstlisting}[numbers=none]
$count[work(adam) $== 8, work(smith) $== 8,
       work(lea)  $== 8, work(john)  $== 8] $== fulltime.
\end{lstlisting}
It constrains the number of variables in $\{$\cvar{work(adam)},
\cvar{work(smith)}, \cvar{work(lea)}, \cvar{work(john)}$\}$ that are equal to $8$,
to be equal to \cvar{fulltime}.
Constraint variable \cvar{fulltime} counts how many persons are working full time.
Global constraints do have a similar syntax to propositional aggregates.
Also their semantics is similar to \emph{count} aggregates in ASP (cf.~\cite{MPG:M:3.7.2}).
But global constraints only constrain the values of constraint variables,
not propositional ones.

\emph{Clingcon} also supports the global constraint $distinct$,
where
\begin{lstlisting}[numbers=none]
$distinct{work(A) : person(A)}.
\end{lstlisting}
means that all persons should have a different workload.
That is,
all values assigned to constraint variables in $\{$\cvar{work(adam)},
\cvar{work(smith)}, \cvar{work(lea)}, \cvar{work(john)}$\}$
have to be distinct from each other.
This constraint could also be expressed using a quadratic number
of inequalities.
Using a single dedicated constraint is usually much more efficient in terms of
memory consumption and runtime.

As global constraints are usually not supported in a negated form in a CP solver,
we have the syntactic restriction that all global constraints must become
facts during grounding and therefore may only occur in the head of rules.
Further global constraints can easily be integrated into this
generic framework.

A valid solution to our constraint logic program in Listing~\ref{listing:example1} contains
the regular literal \Tmylit{team(adam,smith)},
but also constraint literals like \Tcon{work(lea)\$==0}, \Tcon{work(john)\$==0} and 
\Tcon{work(adam)\$+work(smith)\$>6}.
The solution also contains assignments to constraint variables, like
\cvar{work(adam)}$\mapsto$\cvar{8}, \cvar{work(smith)}$\mapsto$\cvar{3}, \cvar{work(lea)}$\mapsto$\cvar{0}, \cvar{work(john)}$\mapsto$\cvar{0}
and \cvar{fulltime}$\mapsto$\cvar{1}.

\myparagraph{Optimization}
In Line~\ref{line:optimize} we give a maximize statement over constraint variables.
This is also a new feature of {\clingcon}.
We maximize the sum over a set of variables and/or expressions.
In this case, we try to maximize \con{work(adam) \$+ work(smith) \$+ work(lea) \$+ work(john)}.
For optimization statements over constraint variables,
we also rely on the syntax of \textit{gringo's}
propositional optimization statements.
We support minimization/maximization and multi-level optimization.
To distinguish propositional and constraint statements,
we precede the latter with a \texttt{\$} sign.
One optimal solution to the problem contains the propositional literal
\Tmylit{team(adam,john)},
and the constraint literals \Tcon{work(lea)\$==0}, 
\Tcon{work(smith)\$==0},
and \Tcon{work(adam)\$+work(john)\$>6}.
To maximize work load,
the constraint variables are assigned 
\cvar{work(adam)}$\mapsto$\cvar{10}, \cvar{work(smith)}$\mapsto$\cvar{0}, \cvar{work(lea)}$\mapsto$\cvar{0}, \cvar{work(john)}$\mapsto$\cvar{10}
and \cvar{fulltime}$\mapsto$\cvar{0}.

To find a constraint optimal solution,
we have to combine the enumeration techniques of {\clasp}
with the ones from the CP solver.
Therefore,
when we first encounter a full propositional assignment,
we search for an optimal (w.r.t. to the optimize statement)
assignment of the constraint variables using the 
search engine of the CP solver.
Let us explain this with the following constraint logic program.
\begin{lstlisting}[numbers=none]
$domain(1..100).
a :- x $* x $< 25.
$minimize{x}.
\end{lstlisting}
Assume {\clasp} has computed the full assignment $\{\Fcon{x \$* x \$< 25},\Flit{a}\}$.
Afterwards, 
we search for the constraint optimal solution
to the constraint variable \cvar{x} which yields $\{\cvar{x}\mapsto 5\}$.
Given this optimal assignment,
a constraint can be added to the CP solver
that all further solutions
shall be below/above this optimum (\con{x\$<5}).
This constraint will now restrict all further solutions
to be ``better''.
We enumerate further solutions,
using the enumeration techniques of \clasp.
So the next assignment is $\{\Tcon{x \$* x \$< 25},\Tlit{a}\}$
and the CP solver finds the optimal constraint variable assignment
$\{\cvar{x}\mapsto 1\}$.
Each new solution restricts the set of further solutions,
so our constraint is changed to (\con{x\$<1}) which then
allows no further solutions to be found.


\section{Conflict Filtering in \clingcon}\label{sec:minimization}

The development of Conflict Driven Clause Learning (CDCL) algorithms
was a major breakthrough in the area of SAT.
Also, CDCL is crucial in SMT solving (cf.~\cite{niolti06a}).
A prerequisite to combine a CDCL-based SAT solver with a theory solver
is the possibility to generate good conflicts and reasons originating in 
the underlying theory.
Therefore,
modern SMT solvers use their own specialized theory propagators
that can produce such witnesses.
\emph{Clingcon} instead uses a black-box approach regarding theory solving.
In fact, off-the-shelf CP solvers, like \gecode,
do usually not provide any reason for their underlying inference.
As a consequence,
conflict and reason information was so far only crudely approximated in \clingcon.
We address this shortcoming by developing mechanisms for extracting minimal
reasons and conflicts from any CP solver using monotone propagators.
We assume that the reader has basic knowledge on CDCL-based ASP solving,
and direct the interested reader to~\cite{gekanesc07a}.

Whenever the CP solver finds out that the set of constraints is inconsistent
under the current assignment $A$,
a conflicting nogood $N$ must be generated,
which can then be used by the ASP solver in its conflict analysis.
The \emph{simple} version of generating the conflicting nogood $N$,
is just to take the entire assignment of constraint literals.
In this way,
all yet decided constraint atoms constitute 
$N=\{\ell \mid \ell \in \project{A}{\catoms}\}$.
In this case, the corresponding list of inconsistent constraints is
\begin{equation}
  I = [\;\gamma(\ell)\mid \ell \in \project{A}{\catoms}\;]\label{eq:I}.
\end{equation}
In order to reduce this list of inconsistent constraints and to find the real cause of the conflict,
we apply an \emph{Irreducible Inconsistent Set} (IIS) algorithm.
The term IIS was coined in~\cite{loon81} for describing inconsistent sets of constraints
having consistent subsets only.
We use the concept of an IIS to find the minimal cause of a conflict.
With this technique,
it is actually possible to drastically reduce such exhaustive sets of
inconsistent constraints as in~\eqref{eq:I} and to create a much smaller
conflict nogood.
It is now possible to apply an IIS algorithm to every conflicting set of constraints
in order to provide {\clasp} with smaller nogoods.
This enhances the information content of the learnt nogood
and hopefully speeds up the search process by better pruning the search space.
\emph{Clingcon} now features several alternatives to reduce such conflicts.
To this end,
we build upon the approach of~\cite{chidra91} who propose different algorithms
for computing IISs,
among them the so-called \emph{Deletion Filtering} algorithm.
In what follows,
we first present the original idea of \emph{Deletion Filtering}
and afterwards propose several refinements that can then be used
to reduce inconsistent lists of constraints in the context of ASP modulo CSP.

\myparagraph{Deletion Filtering}
Given an inconsistent list of constraints $I=[c_1,\dots,c_n]$ as in (\ref{eq:I})
the \emph{Deletion Filtering} Algorithm~\ref{algo:deletionIIS} reduces it to an irreducible list.
We test for each $c_i\in I$ whether $I\setminus c_i$ is inconsistent or not.
If it is inconsistent we can restart the whole algorithm with the list $I\setminus c_i$
continuing with the next $i$.
\begin{algorithm}[t]
				\Input{An inconsistent list of constraints $I=[c_1,\dots,c_n]$.}
\Output{An irreducible inconsistent list of constraints.}
\BlankLine
$i \leftarrow 1$\label{algo:deletion:init}\;
\While{$i \leq |I|$}
{
		\If{$I\setminus c_i$ is inconsistent\label{algo:deletion:testinconsistent}}
		{
		$I \leftarrow I\setminus c_i$\label{algo:deletion:minus}\;
		}
		\Else
		{
		   $i \leftarrow i+1$\;
		}
}
\Return{I}\;
\caption{\textsc{deletion\_filtering}\label{algo:deletionIIS}}
\end{algorithm}%

The result of this simple approach is a minimal inconsistent list,
as we can see in the following example.
Suppose we branch on \Tmylit{team(adam,lea)}.
Unit propagation implies the literals \Tcon{work(lea)\$==work(adam)}, \Tcon{work(john)\$==0},
\Tcon{work(smith)\$==0}, and \Tcon{work(adam)\$+work(lea)\$>6}.
At this point we cannot do any constraint propagation\footnote{w.l.o.g. we assume arc consistency~\cite{mohe86a}}
and make another choice,
\Tmylit{friday}, and some unit propagation,
resulting in \Tcon{work(lea)\$-work(adam)\$==1}.
As unit propagation is at fixpoint,
the CP solver checks the constraints in the partial assignment $\project{A}{\catoms}$
for consistency.
As it is inconsistent, a \emph{simple} conflicting nogood would be
$N=\{\ell\mid \ell \in \project{A}{\catoms}\}$.
To minimize this nogood,
we now apply \emph{Deletion Filtering} to the list $I$ as defined in (\ref{eq:I}):
\begin{align*}
I&=[\gamma(\ell) \mid \ell \in \project{A}{\catoms}]\\
&=[work(lea)=work(adam),work(john)=0,work(smith)=0]\\
&\hspace{2pt}\circ\hspace{2pt}[work(adam)+work(lea)>6,work(lea)-work(adam)=1]
\end{align*}
For $i=1$, we test
$[work(john)=0, work(smith)=0, work(adam)+work(lea)>6,
work(lea)-work(adam)=1]$,
but it does not lead to inconsistency (Line~\ref{algo:deletion:testinconsistent}).
Next list to test is $[work(lea)=work(adam), work(smith)=0,
work(adam)+work(lea)>6, work(lea)-work(adam)=1]$
which restricts the domains of \cvar{work(adam)} and \cvar{work(lea)} to $\emptyset$.
As this is inconsistent,
we remove $work(john)=0$ from $I$ and go on,
also removing $work(smith)=0$ and $work(adam)+work(lea)>6$.
We end up with the irreducible list
$I=[work(lea)=work(adam), work(lea)-work(adam)=1]$,
and can now build a much smaller conflicting nogood
$N=
\{\gamma^{-1}(c)\mid c \in I\}
=
\{\Tcon{work(lea)\$==work(adam)},\Tcon{work(lea)\$-work(adam)\$==1}\}$
as this really describes the cause of the inconsistency.

In most CP solvers, propagation is done in a constraint space.
This space contains the constraints and the variables of the problem.
After doing propagation,
the domains of the variables are restricted.
Normally in CP solvers like {\gecode} this effect
cannot be undone.
As long as we add further constraints to the constraint space
this is no problem,
as another constraint restricts the domain of the variables even more.
If we want to remove a constraint from a constraint space
we have to create a new space containing only the constraints we want to apply.
Then we have to redo all the propagation.
This is why we identified Line~\ref{algo:deletion:minus} in
Algorithm~\ref{algo:deletionIIS} as an efficiency bottleneck.
To address this problem,
we propose some derivatives of the algorithm.
\begin{algorithm}[t]
				\Input{An inconsistent list of constraints $I=[c_1,\dots,c_n]$.}
\Output{An irreducible inconsistent list of constraints $I'$.}
\BlankLine
$I' \leftarrow []$\label{algo:forward:init}\;
\While{$I'$ is consistent\label{algo:forward:return}}
{
  $T \leftarrow I'$\label{algo:forward:reinit}\;
	$i \leftarrow 1$\;
	\While{$T$ is consistent\label{algo:forward:testinconsistent}}
	{
	$T \leftarrow T \circ c_i$\;\label{algo:forward:extend}
	$i \leftarrow i + 1$\;
	}
  $I' \leftarrow I' \circ c_i$\label{algo:forward:extendresult}\;
}
\Return{$I'$}
\caption{\textsc{forward\_filtering}\label{algo:forwardIIS}}
\end{algorithm}%

%
%
%
\myparagraph{Forward Filtering}
Algorithm~\ref{algo:forwardIIS} is designed to avoid
resetting the search space of the CP solver.
It incrementally adds constraints to a testing list $T$,
starting from the first assigned constraint to the last one (lines \ref{algo:forward:testinconsistent} and \ref{algo:forward:extend}).
Remember that incrementally adding constraints is easy for a CP solver
as it can only further restrict the domains.
%
%
If our test list $T$ becomes inconsistent 
we add
the currently tested constraint to the result $I'$ (lines \ref{algo:forward:testinconsistent} and \ref{algo:forward:extendresult}).
If this result is inconsistent (Line \ref{algo:forward:return}), we have found a minimal
list of inconsistent constraints.
Otherwise, we start again,
this time adding all yet found constraints $I'$ to our testing list $T$
(Line \ref{algo:forward:init}).
Now we have to create a new constraint space.
But by incrementally increasing the testing list,
we already reduced the number of potential candidates that
contribute to the IIS,
as we never have to check a constraint beyond the last added constraint.
We illustrate this again on our example.
We start Algorithm~\ref{algo:forwardIIS} with $T=I'=[]$ and 
\begin{align*}
I&=[work(lea)=work(adam),work(john)=0,work(smith)=0]\\
&\hspace{2pt}\circ\hspace{2pt}[work(adam)+work(lea)>6,work(lea)-work(adam)=1]
\end{align*}
in Line \ref{algo:forward:reinit}.
We add $work(lea)=work(adam)$ to $T$,
as this constraint alone is consistent,
we loop and add constraints until $T=I$.
As this list is inconsistent,
we add the last constraint $work(lea)-work(adam)=1$ to $I'$ in Line \ref{algo:forward:extendresult}.
We can do so,
as we know that the last constraint is indispensable for the inconsistency.
As $I'$ is consistent we restart the whole procedure,
but this time setting $T=I'=[work(lea)-work(adam)=1]$ in Line \ref{algo:forward:reinit}.
Please note that,
even if $I$ would contain further constraints,
we would never have to check a constraint behind
$work(lea)-work(adam)=1$.
Our testing list already contained an inconsistent set
of constraints,
consequently we can restrict ourself to this subset.
Now we start the loop again,
adding $work(lea)=work(adam)$ to $T$.
On their own,
these two constraints are inconsistent,
as there exists no valid pair of values for the variables.
So we add $work(lea)=work(adam)$ to $I'$,
resulting in $I'=[work(lea)-work(adam)=1,work(lea)=work(adam)]$.
This is then our reduced list of constraints and
the same IIS as we got with the \emph{Deletion Filtering} method
(as it is the only IIS of the example).
But this time we only needed one reset of the constraint space (Line~\ref{algo:forward:reinit})
instead of five.

\myparagraph{Backward Filtering}
The basic idea of this algorithm is the same as in Algorithm \ref{algo:forwardIIS}.
But this time, we reverse the order of the inconsistent constraint list.
Therefore,
we first test the last assigned constraint and iterate to the first.
In this way we want to accommodate the fact,
that one of the literals that was decided on the current decision level
has to be included in the conflicting nogood.
Otherwise we would have recognized the conflict before.

\begin{algorithm}[t]
\Input{An inconsistent list of constraints $I=[c_1,\dots,c_n]$.}
\Output{A (possibly smaller) inconsistent list of constraints $I'$.}
\BlankLine
$I' \leftarrow []$\label{algo:range:init}\;
$i\leftarrow n$\;
\While{$I'$ is consistent\label{algo:range:for}}
{
	$I' \leftarrow I' \circ c_i$\label{algo:range:extendtest}\;
	$i \leftarrow i - 1$\;
}
\Return{$I'$}\label{algo:range:return}\;
\caption{\textsc{range\_filtering}\label{algo:rangeIIS}}
\end{algorithm}%

\myparagraph{Range Filtering}
This algorithm does not aim at computing an irreducible list of constraints,
but tries to approximate a smaller one
to find a nice tradeoff between reduction 
of size and runtime of the algorithm.
Therefore,
as shown in Algorithm \ref{algo:rangeIIS},
we move through the reversed list of constraints $I$
and add constraints to the result $I'$ until it becomes inconsistent.
In our example we cannot reduce the inconsistent list anymore,
as the first and the last constraint is needed in the IIS.

\begin{algorithm}[t]
\Input{An inconsistent list of constraints $I=[c_1,\dots,c_n]$.}
\Output{An irreducible inconsistent list of constraints $I'$.}
\BlankLine
\lIf{$size(I)=0$}{\Return{$\emptyset$}}\;
$I' \leftarrow []$\label{algo:cc:initbegin}\;
$T \leftarrow []$\;
$\omega \leftarrow vars(c_n)$\label{algo:cc:initchecker}\label{algo:cc:initend}\;
\While{$I'$ is consistent\label{algo:cc:checkreturn}}
{
  $count \leftarrow |\omega|$\label{algo:cc:count}\;
	$i \leftarrow size(I)$\;
	\While{$T$ is consistent and $i\geq 0$\label{algo:cc:for}}
  {
					\If{$\omega \cap vars(c_i) \neq \emptyset$\label{algo:cc:vartest}}
					{
						$T \leftarrow T \circ c_i$\;
						$\omega \leftarrow \omega \cup vars(c_i)$\;
					
					}
					$i \leftarrow i - 1$\;
	}
  \uIf{$T$ is inconsistent\label{algo:cc:test}}
  {
		$I' \leftarrow I' \circ c_i$\label{algo:cc:add}\;
		$\omega \leftarrow \{y\mid y \in vars(x)$ where $x$ in $I'\}$\;
		$I \leftarrow remove(T,c_i)$\label{algo:cc:remove}\;
		$T \leftarrow I'$\;
	}
	\lIf{$count=|\omega|$}{$\omega \leftarrow \{y\mid y \in vars(x)$ where $x$ in $I\}$\label{algo:cc:fallback}}
}
\Return{$I'$}\label{algo:cc:return}
\caption{\textsc{connected\_component\_filtering}\label{algo:ccIIS}}
\end{algorithm}%

%
%
%
%
%

\myparagraph{Connected Components Filtering}
Algorithm \ref{algo:ccIIS} tries to make use of the structure of the constraints.
Therefore,
it does not go forward or backward through the list of constraints but follows
their used constraint variables.
We start with initializing our result $I'$ and the test list $T$ and set our set 
of observed constraint variables $\omega$ to the variables inside the last assigned constraint
of our list $I$ (lines \ref{algo:cc:initbegin} to \ref{algo:cc:initend}).
Then we start our main loop, remembering how many variables we have seen so far (Line \ref{algo:cc:count}). 
We go over our reversed list of constraints $I$ (Line \ref{algo:cc:for}).
If we find a constraint that contains some of the already inspected variables (Line \ref{algo:cc:vartest}),
we add it to our testing list $T$ and extend the set of already seen variables $\omega$.
We then continue iterating until our testing list $T$ becomes inconsistent (Line \ref{algo:cc:test}).
In this case we add the last tested constraint to our result list $I'$ (Line~\ref{algo:cc:add}).
If this list is already inconsistent (Line~\ref{algo:cc:checkreturn}),
we return a minimal list of constraints (Line~\ref{algo:cc:return}).
If not, we restart the loop.
But this time the set of already seen variables is restricted to the set of variables
in the constraints in $I'$.
Furthermore,
we only iterate over the constraints from the test list (Line \ref{algo:cc:remove}),
as this possibly shorter list is also an inconsistent list of constraints.
If at one point we have not found any non tested constraint that has common variables with the tested ones
(this can be the case if the last constraint of our input list $I$ is not contained in the minimal list of constraints),
we simply add all variables to $\omega$ in Line~\ref{algo:cc:fallback}
so that we do not miss any constraint in the next iteration.
With this algorithm,
we want to take account of the internal structure of the constraints.
For our example, this means we start with the last constraint $work(lea)-work(adam)=1$
and completely ignore $work(john)=0$ and $work(smith)=0$ as their variables
do not occur anywhere in the other constraints.
We end up with the same IIS as with \emph{Forward Filtering}
without checking all constraints that do not have common variables
with the constraints from the IIS.

\myparagraph{Connected Components Range Filtering}
This algorithm is a combination of the \emph{Connected Components Filtering}
and the \emph{Range Filtering} algorithms.
That is why it does not compute an irreducible list of constraints.
We move through the list $I$ like in Algorithm \ref{algo:ccIIS}
and once our test list $T$ becomes inconsistent we simply return it.
This shall combine the advantages of using the structure of the constraints
in the \emph{Connected Components Filtering} and the simplicity 
of the \emph{Range Filtering}.
We ignore $work(john)=0$ and $work(smith)=0$ and end up
with $I'=\{work(lea)-work(adam)=1,work(adam)+work(lea)>6,work(lea)=work(adam)\}$.

\section{Reason Filtering in \clingcon}
Up to now we only considered reducing an inconsistent list 
of constraints to reduce the size of a conflicting nogood.
But we can do even more.
If the CP solver propagates the literal $l$,
a \emph{simple} reason nogood is $N=\{\ell\mid \ell \in \project{A}{\catoms}\}\cup\{\inv{l}\}$.
If we have for example $\project{A}{\catoms}=\{\Tcon{work(john)\$==0,\Tcon{work(lea)-work(adam)\$==1}}\}$,
the CP solver propagates the literal \Fcon{work(lea)\$==work(adam)}.
To use the proposed algorithms to reduce a reason nogood we first
have to create an inconsistent list of constraints.
As $J = [\gamma(\ell)\mid \ell \in \project{A}{\catoms}]$ implies $\gamma(l)$,
this inconsistent list is $I=J\circ[\inv{\gamma(l)}]=[work(john)=0,{work(lea)-work(adam)=1},work(lea)= work(adam)]$.
%
So we can now use these various filtering methods also to reduce
reasons generated by the CP solver.
In this case the reduced reason is $\{\Tcon{work(lea)-work(adam)\$==1},\Tcon{work(lea)\$==work(adam)}\}$.
Smaller reasons reduce the size of conflicts
even more,
as they are constructed using unit resolution.

These two new features of {\clingcon} 
are available via the command line parameters:
\texttt{--csp-reduce-conflict=X} and \texttt{--csp-reduce-reason=X}
where $\texttt{X}=\{\texttt{simple,forward,backward,range,cc,ccrange}\}$.
%
%

\section{The {\clingcon} System}\label{sec:system}
The new filtering methods enhance the learning capabilities of \clingcon.
However,
the new version also features \emph{Initial Lookahead},
\emph{Optimization} and \emph{Propagation Delay}.
We will now present these in more detail.
\myparagraph{Initial Lookahead}
As shown in~\cite{yuma06a},
initial lookahead on constraints can be very helpful
in the context of SMT.
It makes implicit knowledge (stored in the propagators of the theory solver)
explicitly available to the propositional solver.
Our \emph{Initial Lookahead},
which can be enabled using the option
\texttt{--csp-initial-lookahead=<true/false>},
is restricted to constraint literals.
As a preprocessing step,
all of them are separately set to true
and constraint propagation is done.
In this way, binary relations between constraints
become explicitly available to the ASP solver.
For example,
\Tcon{work(smith)\$==0} implies \Fcon{work(smith)-work(adam)\$==1}
whereas
\Tcon{work(lea)\$==work(adam)} implies \Fcon{work(lea)-work(adam)\$==1}.
These are then directly translated into a nogood.
Or more formal:
all constraints $c$ implied by a constraint literal $\Tlit{\ell}$
w.r.t. the theory are added to {\clasp} as the respective binary nogood $\{\Tlit{\ell},\inv{\gamma^{-1}(c)}\}$.
\myparagraph{Optimization}
As shown in Section~\ref{sec:language},
{\clingcon} now supports optimization statements over constraint
variables.
The option \texttt{--csp-opt-val}
expects a comma-separated list of values,
similar to the \textit{clasp} 1.3 option \texttt{--opt-val}.
With this option a value for every constraint optimization statement
can be given.
The solver will then start the search using these values.
This is especially useful in combination with the option
\texttt{--csp-opt-all},
that is used to compute all models that are less
or equal to the last found bound.
It forms the logical equivalent to the \textit{clasp} 1.3 option \texttt{--opt-all}.
To compute all constraint-optimal solutions,
one first computes one optimal solution.
Afterwards,
given the optimum value,
the same encoding can be used with the options \texttt{--csp-opt-val}
and \texttt{--csp-opt-all} to compute all optimal solutions.
%
\myparagraph{Propagation Delay}
With \emph{Propagation Delay} we have a new experimental feature
that balances the interplay between the ASP and the CSP part.
Constraint propagation can be expensive,
especially in combination with the filtering techniques from
Section~\ref{sec:minimization}.
It might be beneficial to give more attention to the ASP solver.
This can be done by skipping constraint propagations.
Whenever we can propagate a constraint atom or encounter a conflict
with the CP solver,
filtering methods can be applied.
If we therefore skip constraint propagation and only do it every
$n$'th time,
{\clasp} has the chance to find more conflicts.
If we learn less from the CSP side, we learn more from the ASP side.
The option \texttt{--csp-prop-delay=n} where $n\in \mathds{N}^+_0$ 
can be used to set the propagation delay:
\begin{itemize}
\item $n=1$ does constraint propagation every time, similar to the old \clingcon,
\item $n>1$ does constraint propagation only every $n$'th time and
\item $n=0$ does constraint propagation only on a full propositional assignment.
\end{itemize}
Whenever we do constraint propagation we have to catch up on the
missed propagation.
%
%
%

\section{Experiments}\label{sec:experiments}

We collected various benchmarks from different
categories to evaluate the effects of our new features
on a broad range of problems.
All of them can be expressed using
a mixed representation of Boolean and non-Boolean variables.
We restrict ourself to classes where the ASP and the CSP
part interact tightly to solve the problem,
as we focus on the learning capabilities between
the two systems.
For encodings where we do not have an ASP or a CSP part,
we will not see any effect of our new features.
We now present our benchmark classes with a short description
of the used encodings.
All encodings and instances can be found online at~\cite{clingcon}.

\myparagraph{Benchmarks}
Given a set of squares, the \emph{Packing} problem is to pack all squares
into a rectangular area with fixed dimension.
This problem is directly taken from the 2011 ASP Competition~\cite{ASPcomp11}.
The position of the corners of the squares can be represented using integer variables.
and are guessed by the CP solver.
Within ASP we only have to check whether two squares overlap.

\emph{Incremental Scheduling} is a problem variant
of the well-known job shop scheduling problem which requires rescheduling
and ordering of jobs.
Furthermore, all jobs have a deadline,
and,
if a job finishes after its deadline,
the difference is taken as ``tardiness'' of the job.
This tardiness multiplied by the importance of the job results in a penalty.
The task is to find a solution where the sum of all penalties is below a
given maximum.
The problem is also taken from the 2011 ASP Competition.
We use constraint variables to denote the starting times of the jobs and
also to compute the tardiness and penalties.

Taken from the 2011 ASP Competition,
the \emph{Weighted Tree} problem is inspired by cost-based ``join-order''
optimization of SQL queries in databases.
The problem is to find a full binary tree with $n$ leaves such that
its leaves are pairs (weight, cardinality) of integers,
the right child of an inner node is a leaf,
where its color is either green, red, or blue, and
there are $1,\dots,n-1$ inner nodes such that node $n-1$ is a root node and inner node $i-1$
is the left child of inner node $i$ for $i=2,\dots,n-1$.
The weights of inner nodes are computed recursively based on their colors,
and the weights and cardinalities of their children.
We use constraint variables to represent the cardinality and the weight of the nodes.
We furthermore extended the problem to an optimization problem that tries to
find the ``cheapest'' tree by minimizing the sum of the leaf weights according to
the structure of the tree. 
For this we use the new optimization statement over constraint variables.
In our benchmarks, an instance is solved if we have found and proven the optimal solution.

Given an $n\times n$ board,
placing numbers in the range $\{1,\dots, n\}$ such that there
are no two equal numbers in the same row/column
is called \emph{Quasi Group} problem.
For our benchmarks,
we let $n=20$.
We assign random numbers to $0-80$ percent
of the fields.

To increase the spectrum of benchmarks,
we conceived a new collection of benchmarks which make use of the CP solver
to do the \emph{Unfounded Set Check (USC)}
for some normal logic programs.
Therefore, we reify~\cite{gekasc11a,potasscoManual} logic programs
(in our case we take \emph{Labyrinth} -- the problem of guiding an avatar through a dynamically changing labyrinth to certain fields~\cite{ASPcomp11},
\emph{HashiwoKakero} -- a logic puzzle game and \emph{HamiltonianCycle}).
Using this reified program we can reason about the structure of the program.
In particular, we can add an encoding that does the unfounded set check
using level-mapping as proposed in~\cite{niemela08a}.
We assign a level to every atom in a strongly connected component
and use the CP solver to find a valid mapping.
Using this translation,
we can solve any non-tight logic program using
the CP solver for the unfounded set check.

\myparagraph{Settings}
We run our benchmarks single-threaded on a cluster
with $24\times8$ cores with 2.27GHz each.
We restricted each run to use 4GB RAM.
In all our benchmarks we used the standard configuration of
\clingcon, unless stated otherwise.
We now evaluate the new features of \clingcon.

\myparagraph{Global Constraints}
We want to check whether the use of global constraints
speeds up the computation.
Therefore we have chosen the \emph{Quasi Group} problem,
as it can be easily expressed using the global constraint \emph{distinct}.
We compare two different encodings for \emph{Quasi Group}.
The first one uses one \emph{distinct} constraint for every row and every column.
The second one uses a cubic number of inequality constraints.
We tested 78 randomly generated instances of size $20\times 20$.
While the first encoding using the global constraints results in
an average runtime of 220 seconds and 18 timeouts over all instances,
the decomposed version was much slower.
It used 391 seconds on average and had 27 timeouts.
\emph{Clingcon} confirms,
that global constraints are handled more efficiently
than their explicit decomposition.

\myparagraph{Initial Lookahead}
In Section~\ref{sec:system},
we presented \emph{Initial Lookahead}
over constraints as a new feature of \clingcon.
We now want to study in which cases this technique can be useful 
in terms of runtime.
We run all our benchmarks once with and without \emph{Initial Lookahead}.
In Table~\ref{tab:prepro},
the first column shows the problem class and its number of instances.
The second and the third column show the average runtime in seconds
that is used with and without \emph{Initial Lookahead (I.L.)}.
Timeouts are shown in parenthesis.
The last two columns show the average runtime of the lookahead
algorithm and the number of nogoods that have been produced on
average per instance.
As we can see for the problems \emph{Packing}, \emph{Quasi Group},
and \emph{Weighted Tree},
direct relations between constraints
can be detected and the overall runtime can therefore be reduced.
But this technique does not work on all benchmark classes.
For \emph{Incremental Scheduling} relations between constraints are found
but the additional nogoods seem to deteriorate the performance of the solver.
In the case of the \emph{Unfounded Set Check},
nearly no relations have been found,
so no difference in runtime can be detected.
%
\begin{table}
        \center
        \begin{tabular}{| l | r | r | r | r |}
                \cline{1-5}
                instances                           & time      & time        & time   & nogoods\\
                (\#number)                          & (timeouts)& with \textit{I.L }   & of \textit{I.L.}& from \textit{I.L}\\
                \cline{1-5}
								\emph{Packing}(50)                  & 888(49)   & 882(49)   & 5      & 7970  \\
								\emph{Inc. Sched.}(50)              &  30(01)   &  40(02)   & 0      &   73  \\
								\emph{Quasi Group}(78)              & 390(28)   & 355(24)   & 9      &105367 \\
								\emph{Weighted Tree}(30)            & 484(07)   & 312(04)   & 0      & 1520 \\
								\emph{USC}(132)                    & 721(104)  & 719(103)  & 3      & 1 \\
                \cline{1-5}
        \end{tabular}
        \caption{Initial Lookahead (\textit{I.L.})}
        \label{tab:prepro}
\end{table}

\myparagraph{Conflict and Reason Filtering}
We want to analyze how much the different conflict and reason filtering methods
presented in Section~\ref{sec:minimization}
differ in size of conflicts and average runtime.
As conflicts and reasons are strongly interacting in the CDCL framework,
we test the combination of all our proposed algorithms.
We denote the filtering algorithms with the following shortcuts:
s(\emph{Simple}), b(\emph{Backward Filtering}), f(\emph{Forward Filtering}), c(\emph{Connected Component Filtering}),
r(\emph{Range Filtering}) and o(\emph{Connected Component Range Filtering}).
We name the filtering algorithm for reasons first,
separated by a slash from the algorithm used to filter conflicts.
To denote the configuration using \emph{Range Filtering} for reasons
and \emph{Forward Filtering} for conflicts,
we simply write r/f.
The original configuration,
which can be seen as the ``old'' {\clingcon} is therefore denoted s/s.
We start by showing the impact on average conflict size of all configurations
using a heat map in Figure~\ref{figure:acl1}.
It shows the reduction of the conflict size in percentage
relative to the worst configuration.
The rows represent the used algorithms for reason filtering,
the columns represent the algorithms for filtering conflicts.
So the worst configuration is represented by a totally black square
and a configuration that reduces the average conflict size by half is gray.
A completely white field would mean that the conflict size has been reduced to zero.
\newcounter{col}
\newcounter{row}
\newcounter{result}
\newlength{\rowl}
\newlength{\coll}
%
%
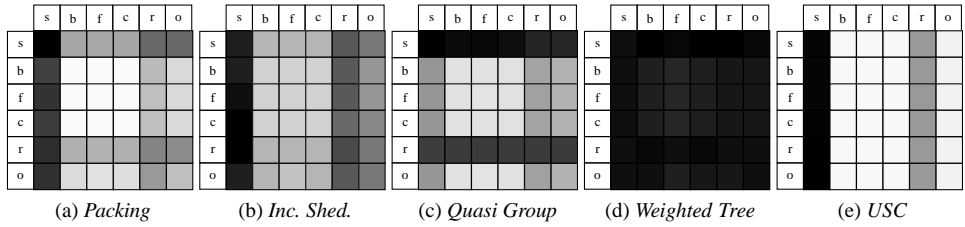
\begin{figure}
				\subfloat[\emph{Packing}]{\tikzstyle{square} = [draw=black, minimum width=3.5mm, minimum height=3.5mm]
\begin{tikzpicture}[remember picture]
        \footnotesize
			\node[name=rs] at (-3.5mm,23.0mm-3.50mm) [square] {\tiny{s}};
			\node[name=rb] at (-3.5mm,23.0mm-3.50mm*2) [square] {\tiny{b}};
			\node[name=rf] at (-3.5mm,23.0mm-3.50mm*3) [square] {\tiny{f}};
			\node[name=rc] at (-3.5mm,23.0mm-3.50mm*4) [square] {\tiny{c}};
			\node[name=rr] at (-3.5mm,23.0mm-3.50mm*5) [square] {\tiny{r}};
			\node[name=rcr] at (-3.5mm,23.0mm-3.50mm*6) [square] {\tiny{o}};

			\node[name=cs] at (0mm,23mm) [square] {\tiny{s}};
			\node[name=cb] at (3.50mm,23mm) [square] {\tiny{b}};
			\node[name=cf] at (3.50mm*2,23mm) [square] {\tiny{f}};
			\node[name=cc] at (3.50mm*3,23mm) [square] {\tiny{c}};
			\node[name=cr] at (3.50mm*4,23mm) [square] {\tiny{r}};
			\node[name=ccr] at (3.50mm*5,23mm) [square] {\tiny{o}};

			\foreach \x in {293,104,99,104,173,173,219,5,5,5,79,44,234,6,5,5,72,43,223,5,5,5,73,43,239,92,87,91,142,131,236,40,35,38,118,74}
			{
			\pgfmathparse{round((\x/293)*100)} 

			\setcounter{result}{\pgfmathresult}
      \pgfmathparse{\value{col}*1.24mm}
      \setlength{\coll}{\pgfmathresult mm}
      \pgfmathparse{(1.95mm-\value{row})*1.24mm}
      \setlength{\rowl}{\pgfmathresult mm}
			
			\node[name=s\x] at (\the\coll,\the\rowl) [square, fill=black!\theresult] {{}};
			\ifnum\value{col}=5
						\setcounter{col}{0}
						\addtocounter{row}{1}
		  \else
						\addtocounter{col}{1}
			\fi
			}
			\setcounter{col}{0}
			\setcounter{row}{0}

  \end{tikzpicture}}
        \subfloat[\emph{Inc. Shed.}]{\tikzstyle{square} = [draw=black, minimum width=3.5mm, minimum height=3.5mm]
\begin{tikzpicture}[remember picture]
        \footnotesize
			\node[name=rs] at (-3.5mm,23.0mm-3.50mm) [square] {\tiny{s}};
			\node[name=rb] at (-3.5mm,23.0mm-3.50mm*2) [square] {\tiny{b}};
			\node[name=rf] at (-3.5mm,23.0mm-3.50mm*3) [square] {\tiny{f}};
			\node[name=rc] at (-3.5mm,23.0mm-3.50mm*4) [square] {\tiny{c}};
			\node[name=rr] at (-3.5mm,23.0mm-3.50mm*5) [square] {\tiny{r}};
			\node[name=rcr] at (-3.5mm,23.0mm-3.50mm*6) [square] {\tiny{o}};

			\node[name=cs] at (0mm,23mm) [square] {\tiny{s}};
			\node[name=cb] at (3.50mm,23mm) [square] {\tiny{b}};
			\node[name=cf] at (3.50mm*2,23mm) [square] {\tiny{f}};
			\node[name=cc] at (3.50mm*3,23mm) [square] {\tiny{c}};
			\node[name=cr] at (3.50mm*4,23mm) [square] {\tiny{r}};
			\node[name=ccr] at (3.50mm*5,23mm) [square] {\tiny{o}};

			\foreach \x in {15,5,5,5,11,9,15,3,3,3,11,7,16,3,3,3,11,7,17,3,3,3,10,8,17,5,5,5,12,9,15,5,4,5,11,9}
			{
			\pgfmathparse{round((\x/17)*100)} 
     	\setcounter{result}{\pgfmathresult}
      \pgfmathparse{\value{col}*1.24mm}
      \setlength{\coll}{\pgfmathresult mm}
      \pgfmathparse{(1.95mm-\value{row})*1.24mm}
      \setlength{\rowl}{\pgfmathresult mm}
			
			\node[name=s\x] at (\the\coll,\the\rowl) [square, fill=black!\theresult] {{}};
			\ifnum\value{col}=5
						\setcounter{col}{0}
						\addtocounter{row}{1}
		  \else
						\addtocounter{col}{1}
			\fi
			}
			\setcounter{col}{0}
			\setcounter{row}{0}

  \end{tikzpicture}}
        \subfloat[\emph{Quasi Group}]{\tikzstyle{square} = [draw=black, minimum width=3.5mm, minimum height=3.5mm]
\begin{tikzpicture}[remember picture]
        \footnotesize
			\node[name=rs] at (-3.5mm,23.0mm-3.50mm) [square] {\tiny{s}};
			\node[name=rb] at (-3.5mm,23.0mm-3.50mm*2) [square] {\tiny{b}};
			\node[name=rf] at (-3.5mm,23.0mm-3.50mm*3) [square] {\tiny{f}};
			\node[name=rc] at (-3.5mm,23.0mm-3.50mm*4) [square] {\tiny{c}};
			\node[name=rr] at (-3.5mm,23.0mm-3.50mm*5) [square] {\tiny{r}};
			\node[name=rcr] at (-3.5mm,23.0mm-3.50mm*6) [square] {\tiny{o}};

			\node[name=cs] at (0mm,23mm) [square] {\tiny{s}};
			\node[name=cb] at (3.50mm,23mm) [square] {\tiny{b}};
			\node[name=cf] at (3.50mm*2,23mm) [square] {\tiny{f}};
			\node[name=cc] at (3.50mm*3,23mm) [square] {\tiny{c}};
			\node[name=cr] at (3.50mm*4,23mm) [square] {\tiny{r}};
			\node[name=ccr] at (3.50mm*5,23mm) [square] {\tiny{o}};

			\foreach \x in {480,463,467,452,411,409,197,58,57,59,174,144,198,56,55,57,178,143,197,58,57,59,174,144,364,372,369,368,370,368,198,56,55,57,178,143}
			{
			\pgfmathparse{round((\x/480)*100)} 

			\setcounter{result}{\pgfmathresult}
      \pgfmathparse{\value{col}*1.24mm}
      \setlength{\coll}{\pgfmathresult mm}
      \pgfmathparse{(1.95mm-\value{row})*1.24mm}
      \setlength{\rowl}{\pgfmathresult mm}

			\node[name=s\x] at (\the\coll,\the\rowl) [square, fill=black!\theresult] {{}};
			\ifnum\value{col}=5
						\setcounter{col}{0}
						\addtocounter{row}{1}
		  \else
						\addtocounter{col}{1}
			\fi
			}
			\setcounter{col}{0}
			\setcounter{row}{0}

  \end{tikzpicture}}
				\subfloat[\emph{Weighted Tree}]{\tikzstyle{square} = [draw=black, minimum width=3.5mm, minimum height=3.5mm]
\begin{tikzpicture}[remember picture]
        \footnotesize
			\node[name=rs] at (-3.5mm,23.0mm-3.50mm) [square] {\tiny{s}};
			\node[name=rb] at (-3.5mm,23.0mm-3.50mm*2) [square] {\tiny{b}};
			\node[name=rf] at (-3.5mm,23.0mm-3.50mm*3) [square] {\tiny{f}};
			\node[name=rc] at (-3.5mm,23.0mm-3.50mm*4) [square] {\tiny{c}};
			\node[name=rr] at (-3.5mm,23.0mm-3.50mm*5) [square] {\tiny{r}};
			\node[name=rcr] at (-3.5mm,23.0mm-3.50mm*6) [square] {\tiny{o}};

			\node[name=cs] at (0mm,23mm) [square] {\tiny{s}};
			\node[name=cb] at (3.50mm,23mm) [square] {\tiny{b}};
			\node[name=cf] at (3.50mm*2,23mm) [square] {\tiny{f}};
			\node[name=cc] at (3.50mm*3,23mm) [square] {\tiny{c}};
			\node[name=cr] at (3.50mm*4,23mm) [square] {\tiny{r}};
			\node[name=ccr] at (3.50mm*5,23mm) [square] {\tiny{o}};

			\foreach \x in {31,33,32,33,33,32,31,29,28,29,30,30,31,29,28,29,30,30,31,29,28,29,30,30,31,32,31,32,31,31,31,31,29,30,30,31}
			{
			\pgfmathparse{round((\x/33)*100)} 
			\setcounter{result}{\pgfmathresult}
      \pgfmathparse{\value{col}*1.24mm}
      \setlength{\coll}{\pgfmathresult mm}
      \pgfmathparse{(1.95mm-\value{row})*1.24mm}
      \setlength{\rowl}{\pgfmathresult mm}

			\node[name=s\x] at (\the\coll,\the\rowl) [square, fill=black!\theresult] {{}};
			\ifnum\value{col}=5
						\setcounter{col}{0}
						\addtocounter{row}{1}
		  \else
						\addtocounter{col}{1}
			\fi
			}
			\setcounter{col}{0}
			\setcounter{row}{0}

  \end{tikzpicture}}
				\subfloat[\emph{USC}]{\tikzstyle{square} = [draw=black, minimum width=3.5mm, minimum height=3.5mm]
\begin{tikzpicture}[remember picture]
        \footnotesize
			\node[name=rs] at (-3.5mm,23.0mm-3.50mm) [square] {\tiny{s}};
			\node[name=rb] at (-3.5mm,23.0mm-3.50mm*2) [square] {\tiny{b}};
			\node[name=rf] at (-3.5mm,23.0mm-3.50mm*3) [square] {\tiny{f}};
			\node[name=rc] at (-3.5mm,23.0mm-3.50mm*4) [square] {\tiny{c}};
			\node[name=rr] at (-3.5mm,23.0mm-3.50mm*5) [square] {\tiny{r}};
			\node[name=rcr] at (-3.5mm,23.0mm-3.50mm*6) [square] {\tiny{o}};

			\node[name=cs] at (0mm,23mm) [square] {\tiny{s}};
			\node[name=cb] at (3.50mm,23mm) [square] {\tiny{b}};
			\node[name=cf] at (3.50mm*2,23mm) [square] {\tiny{f}};
			\node[name=cc] at (3.50mm*3,23mm) [square] {\tiny{c}};
			\node[name=cr] at (3.50mm*4,23mm) [square] {\tiny{r}};
			\node[name=ccr] at (3.50mm*5,23mm) [square] {\tiny{o}};

			\foreach \x in {454,13,14,15,188,23,458,13,14,14,186,21,458,13,14,14,187,21,459,13,14,14,188,21,464,13,14,14,187,22,465,13,14,14,188,22}
			{
			\pgfmathparse{round((\x/465)*100)} 
			\setcounter{result}{\pgfmathresult}
      \pgfmathparse{\value{col}*1.24mm}
      \setlength{\coll}{\pgfmathresult mm}
      \pgfmathparse{(1.95mm-\value{row})*1.24mm}
      \setlength{\rowl}{\pgfmathresult mm}

			\node[name=s\x] at (\the\coll,\the\rowl) [square, fill=black!\theresult] {{}};
			\ifnum\value{col}=5
						\setcounter{col}{0}
						\addtocounter{row}{1}
		  \else
						\addtocounter{col}{1}
			\fi
			}
			\setcounter{col}{0}
			\setcounter{row}{0}

  \end{tikzpicture}}

        \caption{Average conflict size}
        \label{figure:acl1}
\end{figure}
As we can see in Figure~\ref{figure:acl1},
the average conflict size is reduced by all combinations of filtering
algorithms.
Furthermore,
we see that the first row and column, respectively,  is usually darker
than the others,
which indicates that filtering either only conflicts or only reasons is not enough.
Also we see that for the \emph{Unfounded Set Check (USC)} the filtering
of reasons does not have any effect.
This is due the encoding of the problem.
As nearly no propagation takes place,
no reasons are computed at all.
The shades on the \emph{Range Filtering} rows/columns (denoted by r)
clearly show that the \emph{Range Filtering} produces larger conflicts.
But this is improved by incorporating structure
to the filtering algorithm using \emph{Connected Component Range Filtering}.
Next, we want to see if the reduction of the average conflict size also pays off
in terms of runtime.
Therefore Figure~\ref{figure:time1} shows the heat map for average 
runtime.
\begin{figure}
				\subfloat[\emph{Packing}]{\tikzstyle{square} = [draw=black, minimum width=3.5mm, minimum height=3.5mm]
\begin{tikzpicture}[remember picture]
        \footnotesize
			\node[name=rs] at (-3.5mm,23.0mm-3.50mm) [square] {\tiny{s}};
			\node[name=rb] at (-3.5mm,23.0mm-3.50mm*2) [square] {\tiny{b}};
			\node[name=rf] at (-3.5mm,23.0mm-3.50mm*3) [square] {\tiny{f}};
			\node[name=rc] at (-3.5mm,23.0mm-3.50mm*4) [square] {\tiny{c}};
			\node[name=rr] at (-3.5mm,23.0mm-3.50mm*5) [square] {\tiny{r}};
			\node[name=rcr] at (-3.5mm,23.0mm-3.50mm*6) [square] {\tiny{o}};

			\node[name=cs] at (0mm,23mm) [square] {\tiny{s}};
			\node[name=cb] at (3.50mm,23mm) [square] {\tiny{b}};
			\node[name=cf] at (3.50mm*2,23mm) [square] {\tiny{f}};
			\node[name=cc] at (3.50mm*3,23mm) [square] {\tiny{c}};
			\node[name=cr] at (3.50mm*4,23mm) [square] {\tiny{r}};
			\node[name=ccr] at (3.50mm*5,23mm) [square] {\tiny{o}};

			\foreach \x in {888,246,293,223,772,565,754,27,33,25,178,50,820,23,32,23,144,67,733,25,31,22,120,51,882,338,343,308,794,614,869,63,85,61,589,141}
			{
			\pgfmathparse{round((\x/888)*100)} 

      \setcounter{result}{\pgfmathresult}
      \pgfmathparse{\value{col}*1.24mm}
      \setlength{\coll}{\pgfmathresult mm}
      \pgfmathparse{(1.95mm-\value{row})*1.24mm}
      \setlength{\rowl}{\pgfmathresult mm}
			
			\node[name=s\x] at (\the\coll,\the\rowl) [square, fill=black!\theresult] {{}};
			\ifnum\value{col}=5
						\setcounter{col}{0}
						\addtocounter{row}{1}
		  \else
						\addtocounter{col}{1}
			\fi
			}
			\setcounter{col}{0}
			\setcounter{row}{0}

  \end{tikzpicture}}
        \subfloat[\emph{Inc. Sched.}]{\tikzstyle{square} = [draw=black, minimum width=3.5mm, minimum height=3.5mm]
\begin{tikzpicture}[remember picture]
        \footnotesize
			\node[name=rs] at (-3.5mm,23.0mm-3.50mm) [square] {\tiny{s}};
			\node[name=rb] at (-3.5mm,23.0mm-3.50mm*2) [square] {\tiny{b}};
			\node[name=rf] at (-3.5mm,23.0mm-3.50mm*3) [square] {\tiny{f}};
			\node[name=rc] at (-3.5mm,23.0mm-3.50mm*4) [square] {\tiny{c}};
			\node[name=rr] at (-3.5mm,23.0mm-3.50mm*5) [square] {\tiny{r}};
			\node[name=rcr] at (-3.5mm,23.0mm-3.50mm*6) [square] {\tiny{o}};

			\node[name=cs] at (0mm,23mm) [square] {\tiny{s}};
			\node[name=cb] at (3.50mm,23mm) [square] {\tiny{b}};
			\node[name=cf] at (3.50mm*2,23mm) [square] {\tiny{f}};
			\node[name=cc] at (3.50mm*3,23mm) [square] {\tiny{c}};
			\node[name=cr] at (3.50mm*4,23mm) [square] {\tiny{r}};
			\node[name=ccr] at (3.50mm*5,23mm) [square] {\tiny{o}};

			\foreach \x in {30,1,2,2,9,2,17,3,5,2,23,10,18,4,5,4,25,7,8,3,4,3,18,7,56,3,3,2,27,10,16,3,2,2,21,5}
			{
			\pgfmathparse{round((\x/56)*100)} 
      \setcounter{result}{\pgfmathresult}
      \pgfmathparse{\value{col}*1.24mm}
      \setlength{\coll}{\pgfmathresult mm}
      \pgfmathparse{(1.95mm-\value{row})*1.24mm}
      \setlength{\rowl}{\pgfmathresult mm}

			\node[name=s\x] at (\the\coll,\the\rowl) [square, fill=black!\theresult] {{}};
			\ifnum\value{col}=5
						\setcounter{col}{0}
						\addtocounter{row}{1}
		  \else
						\addtocounter{col}{1}
			\fi
			}
			\setcounter{col}{0}
			\setcounter{row}{0}

  \end{tikzpicture}}
        \subfloat[\emph{Quasi Group}]{\tikzstyle{square} = [draw=black, minimum width=3.5mm, minimum height=3.5mm]
\begin{tikzpicture}[remember picture]
        \footnotesize
			\node[name=rs] at (-3.5mm,23.0mm-3.50mm) [square] {\tiny{s}};
			\node[name=rb] at (-3.5mm,23.0mm-3.50mm*2) [square] {\tiny{b}};
			\node[name=rf] at (-3.5mm,23.0mm-3.50mm*3) [square] {\tiny{f}};
			\node[name=rc] at (-3.5mm,23.0mm-3.50mm*4) [square] {\tiny{c}};
			\node[name=rr] at (-3.5mm,23.0mm-3.50mm*5) [square] {\tiny{r}};
			\node[name=rcr] at (-3.5mm,23.0mm-3.50mm*6) [square] {\tiny{o}};

			\node[name=cs] at (0mm,23mm) [square] {\tiny{s}};
			\node[name=cb] at (3.50mm,23mm) [square] {\tiny{b}};
			\node[name=cf] at (3.50mm*2,23mm) [square] {\tiny{f}};
			\node[name=cc] at (3.50mm*3,23mm) [square] {\tiny{c}};
			\node[name=cr] at (3.50mm*4,23mm) [square] {\tiny{r}};
			\node[name=ccr] at (3.50mm*5,23mm) [square] {\tiny{o}};

			\foreach \x in {390,540,547,442,613,616,37,37,40,37,37,38,64,54,51,56,81,45,26,23,27,23,27,25,784,776,782,764,789,794,13,12,13,10,15,10}
			{
			\pgfmathparse{round((\x/794)*100)} 

      \setcounter{result}{\pgfmathresult}
      \pgfmathparse{\value{col}*1.24mm}
      \setlength{\coll}{\pgfmathresult mm}
      \pgfmathparse{(1.95mm-\value{row})*1.24mm}
      \setlength{\rowl}{\pgfmathresult mm}
			
			\node[name=s\x] at (\the\coll,\the\rowl) [square, fill=black!\theresult] {{}};
			\ifnum\value{col}=5
						\setcounter{col}{0}
						\addtocounter{row}{1}
		  \else
						\addtocounter{col}{1}
			\fi
			}
			\setcounter{col}{0}
			\setcounter{row}{0}

  \end{tikzpicture}}
				\subfloat[\emph{Weighted Tree}]{\tikzstyle{square} = [draw=black, minimum width=3.5mm, minimum height=3.5mm]
\begin{tikzpicture}[remember picture]
        \footnotesize
			\node[name=rs] at (-3.5mm,23.0mm-3.50mm) [square] {\tiny{s}};
			\node[name=rb] at (-3.5mm,23.0mm-3.50mm*2) [square] {\tiny{b}};
			\node[name=rf] at (-3.5mm,23.0mm-3.50mm*3) [square] {\tiny{f}};
			\node[name=rc] at (-3.5mm,23.0mm-3.50mm*4) [square] {\tiny{c}};
			\node[name=rr] at (-3.5mm,23.0mm-3.50mm*5) [square] {\tiny{r}};
			\node[name=rcr] at (-3.5mm,23.0mm-3.50mm*6) [square] {\tiny{o}};

			\node[name=cs] at (0mm,23mm) [square] {\tiny{s}};
			\node[name=cb] at (3.50mm,23mm) [square] {\tiny{b}};
			\node[name=cf] at (3.50mm*2,23mm) [square] {\tiny{f}};
			\node[name=cc] at (3.50mm*3,23mm) [square] {\tiny{c}};
			\node[name=cr] at (3.50mm*4,23mm) [square] {\tiny{r}};
			\node[name=ccr] at (3.50mm*5,23mm) [square] {\tiny{o}};

			\foreach \x in {484,541,536,553,564,531,581,582,582,583,582,579,583,582,582,584,582,581,579,585,582,583,582,582,555,574,569,578,579,572,554,574,570,577,578,572}
			{
			\pgfmathparse{round((\x/585)*100)} 

      \setcounter{result}{\pgfmathresult}
      \pgfmathparse{\value{col}*1.24mm}
      \setlength{\coll}{\pgfmathresult mm}
      \pgfmathparse{(1.95mm-\value{row})*1.24mm}
      \setlength{\rowl}{\pgfmathresult mm}
			
			\node[name=s\x] at (\the\coll,\the\rowl) [square, fill=black!\theresult] {{}};
			\ifnum\value{col}=5
						\setcounter{col}{0}
						\addtocounter{row}{1}
		  \else
						\addtocounter{col}{1}
			\fi
			}
			\setcounter{col}{0}
			\setcounter{row}{0}

  \end{tikzpicture}}
				\subfloat[\emph{USC}]{\tikzstyle{square} = [draw=black, minimum width=3.5mm, minimum height=3.5mm]
\begin{tikzpicture}[remember picture]
        \footnotesize
			\node[name=rs] at (-3.5mm,23.0mm-3.50mm) [square] {\tiny{s}};
			\node[name=rb] at (-3.5mm,23.0mm-3.50mm*2) [square] {\tiny{b}};
			\node[name=rf] at (-3.5mm,23.0mm-3.50mm*3) [square] {\tiny{f}};
			\node[name=rc] at (-3.5mm,23.0mm-3.50mm*4) [square] {\tiny{c}};
			\node[name=rr] at (-3.5mm,23.0mm-3.50mm*5) [square] {\tiny{r}};
			\node[name=rcr] at (-3.5mm,23.0mm-3.50mm*6) [square] {\tiny{o}};

			\node[name=cs] at (0mm,23mm) [square] {\tiny{s}};
			\node[name=cb] at (3.50mm,23mm) [square] {\tiny{b}};
			\node[name=cf] at (3.50mm*2,23mm) [square] {\tiny{f}};
			\node[name=cc] at (3.50mm*3,23mm) [square] {\tiny{c}};
			\node[name=cr] at (3.50mm*4,23mm) [square] {\tiny{r}};
			\node[name=ccr] at (3.50mm*5,23mm) [square] {\tiny{o}};

			\foreach \x in {721,92,160,98,634,415,703,95,158,93,623,409,707,89,156,94,634,409,708,94,157,92,629,406,715,91,157,96,632,412,712,92,157,94,632,413}
			{
			\pgfmathparse{round((\x/721)*100)} 
      \setcounter{result}{\pgfmathresult}
      \pgfmathparse{\value{col}*1.24mm}
      \setlength{\coll}{\pgfmathresult mm}
      \pgfmathparse{(1.95mm-\value{row})*1.24mm}
      \setlength{\rowl}{\pgfmathresult mm}
			
			\node[name=s\x] at (\the\coll,\the\rowl) [square, fill=black!\theresult] {{}};
			\ifnum\value{col}=5
						\setcounter{col}{0}
						\addtocounter{row}{1}
		  \else
						\addtocounter{col}{1}
			\fi
			}
			\setcounter{col}{0}
			\setcounter{row}{0}

  \end{tikzpicture}}
        \caption{Average time in seconds}
        \label{figure:time1}
\end{figure}
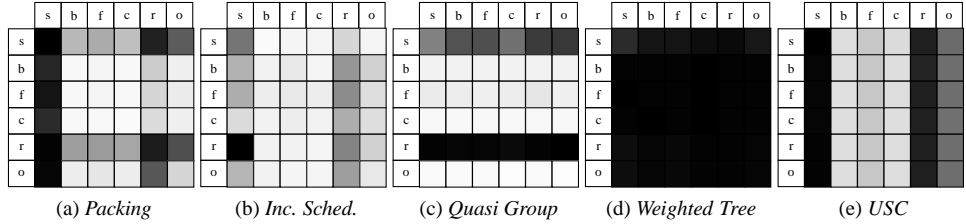
A black square denotes the slowest configuration,
while a gray one is twice as fast.
As we can clearly see,
the reduction of runtime coincides with the reduction of conflict size
in most cases.
Furthermore,
we can see a clear speedup for all benchmark classes
except \emph{Weighted Tree} using the filtering algorithms.
%
\begin{table}
        \center
        \begin{tabular}{| l | r | r | r | r | }
                \cline{1-5}
                Instances                           & time      & time   & acs & acs \\
                (\#number)                          & s/s       & o/b    & s/s & o/b \\
                \cline{1-5}
								\emph{Packing}(50)                  & 888(49) & 63(0)   & 293 & 40 \\
								\emph{Inc. Sched.}(50)   &  30(01) & 3(0)    & 15  & 5 \\
								\emph{Quasi Group}(78)              & 390(28) & 12(0)   & 480 & 56 \\
								\emph{Weighted Tree}(30) & 484(07) & 574(18) & 31  & 31 \\
								\emph{USC}(132)      & 721(104)& 92(1)   & 454 & 13\\
                \cline{1-5}
        \end{tabular}
        \caption{Average time in s(timeouts), average conflict size (acs)}
        \label{tab:delay1}
\end{table}
Table~\ref{tab:delay1} compares the \emph{Simple} version s/s without using
any filtering algorithms,
with the configuration o/b (reducing reasons using \emph{Connected Component Range Filtering}
and reducing conflicts using \emph{Backward Filtering}),
as it has the lowest number of timeouts.
We can see a speedup of around one order of magnitude on all
benchmarks except \emph{Weighted Tree}.
The same picture is given for the reduction of conflict size.
So whenever it is possible to reduce the average conflict size,
this also pays off in terms of runtime.
%
%
%
%
%
\myparagraph{Propagation Delay}
As the filtering of conflicts and reasons takes a lot of time
(e.g configuration o/b uses $43\%$ of the runtime for filtering),
we want to reduce the calls to the filtering algorithms.
Therefore,
with \emph{Propagation Delay},
we can do less propagation with the CP solver and 
it will produce less conflicts and reasons,
hopefully reducing the number of calls.
We therefore take the yet best configuration o/b and compare
different propagation delays
$n\in\{1,10,0\}$ (normal, every ten steps, only on model).
%
%
\begin{table}
\begin{minipage}[b]{0.50\linewidth}\centering
        \begin{tabular}{| l | r | r | r | }
                \cline{1-4}
                 $n$                            &   $1$   & $10$    & $0$ \\
                \cline{1-4}
								\emph{Packing}                  &  31534  &  14897  &  8463 \\
                \emph{Inc. Sched.}              &  3505   &  3240   &  6660 \\
                \emph{Quasi Group}              &  4245   &  1535   &  1726 \\
								\emph{Weighted Tree}            &  6868k  &  1168k  &  1042k\\
								\emph{USC}                      &  2007   &  2118   &  1768\\
                \cline{1-4}
        \end{tabular}
        \caption{Calls to filtering algorithms o/b}
        \label{tab:delaycalls}
\end{minipage}\begin{minipage}[b]{0.50\linewidth}\centering
        \begin{tabular}{| l | r | r | r | }
                \cline{1-4}
                 $n$                            &   $1$   & $10$   & $0$ \\
                \cline{1-4}
								\emph{Packing}                  &  63     &  75    &  571  \\
                \emph{Inc. Sched.}              &  3      &  6     &  11   \\
                \emph{Quasi Group}              &  12     &  9     &  19   \\
								\emph{Weighted Tree}            &  574    &  559   &  546 \\
								\emph{USC}                      &  92     &  91    &  82\\
                \cline{1-4}
        \end{tabular}
        \caption{Times of configuration o/b}
        \label{tab:delaytime}
\end{minipage}
\end{table}
Table~\ref{tab:delaycalls} shows the average number of calls to a filtering algorithm for 
configuration o/b with different delays.
We see a reduction of the number of calls on all benchmarks except on \emph{Incremental Scheduling}
where it doubled.
This is clearly due to a loss of information that is necessary for the search.
If the CP solver has less influence on the search,
the ASP part gets more control.
But the missing knowledge from the CSP part has to be compensated by pure search
in the ASP part.
Therefore, Table~\ref{tab:delaytime} shows that some benchmarks like
\emph{Weighted Tree} and \emph{Unfounded Set Check} can relinquish
some propagation power gaining additional speedup.
On others like \emph{Packing} this propagation is urgently needed to drive the search
and cannot be compensated.
This feature has to be investigated further to gain benefits
for practical usage.

\section{Related Work}
In the quite young field of ASP modulo CSP a lot of research
has been done in the last years.
The approaches can be separated roughly into two classes:
integration and translation.
The integrated approaches like \emph{CASP}~\cite{baboge05a}
and ADSolver/ACSolver~\cite{melgel08a,megezh08a}
are similar to the {\clingcon} system.
However, no learning is used in the approaches,
as the constraint solver just
checks the assignment of constraints.
Later, \cite{balduccini09a} showed how to use ASP
as a specification language,
where each answer set represents a CSP.
In this approach no coupling between the systems
was possible and therefore learning facilities
were not used.
Afterwards GASP~\cite{dadoporo09a} presented a bottom up approach
where the logic program was grounded on the fly.
With Dingo~\cite{jalini11a} ASP was translated to difference logic
using level mapping for the unfounded set check.
An SMT solver is used to solve the translated problem.
Nowadays, more and more translational approaches arise in the area 
of SMT.
\emph{Sugar}~\cite{batata08a} is a very successful solver which translates
the various supported theories to SAT.
Also, in \cite{drewal10a} it was shown how to translate constraints
into ASP during solving.
These translational approaches have the strongest coupling
and therefore the highest learning capabilities.
The reason generation and conflict handling is directly done by the underlying
SAT solver and therefore very efficient.
On the other hand, they do have problems to find a compact
representation of the constraints without losing propagation strength.
\emph{Clingcon} therefore tries to catch up,
improving the learning facilities and still preserving the
advantages of integrated approaches like compact representation
of constraint propagators.
Furthermore,
{\clingcon} is a ASP modulo Theory solver that aims at taking advantage of arbitrary theories
in the long run, eg description logics.
Such a variety can only be supported by a black box approach.
Similar results regarding the filtering methods have been introduced
in~\cite{junker01} but have not been applied to an SMT framework.


\section{Discussion}\label{sec:discussion}
We extended and improve {\clingcon} in various ways.
At first,
the input language was expanded to support \emph{Global Constraints}
and \emph{Optimization Statements} over constraint variables.
As the input language is a big advantage over pure SMT systems,
complex hybrid problems can now easily be expressed as
constraint logic programs.
We have shown that \emph{Initial Lookahead} can give advantages in terms of speedup
on some problems.
We developed \emph{Filtering} methods
for conflicts and reasons that can be applied to any theory solver.
This enables the ASP solver to learn about the structure of the theory,
even if the theory solver does not give any information about it (black box systems).
We furthermore show that while applying these filtering methods,
that knowledge is discovered that is valuable for the overall search process
and can therefore speed up the search by orders of magnitude.
Unfortunately, a direct comparison with existing SMT solvers is inapplicable in view of different input formats.
However, we want to conduct an indirect comparison by translating ASPmCSP problems to SMT
following the line of~\cite{jalini11a}, using a level mapping for the unfounded set check.
This allows us to compare our approach to SMT solvers that do not use
a black-box CP solver but do propagation either with dedicated algorithms
(eg.~\cite{boniol08})
or a translation of CP to SAT (cf.~\cite{batata08a}).
Such solvers have complete control over reason and conflict generation
and can therefore use extra knowledge to create better conflicts.
With \emph{Propagation Delay} we developed a method to control
the impact of the interaction among both systems to the search.
To balance the \emph{Propagation Delay} dynamically during search will be
topic of additional research.
Our work leaves the burden of choosing the right reason and conflict generation strategy
to the user.
Although for our benchmark set the optimal filtering configuration was o/b,
this may vary on other benchmark classes.
This can be counterbalanced by an adaption of the \emph{claspfolio}~\cite{gekakascsczi11a}
system to \emph{clingcon} in order to automatically derive an optimal configuration
of \emph{clingcon} from the features of problem instances.
In the future we still want to focus on learning capacities
and increase the coupling of the two systems,
such that the CP solver also benefits from the
elaborated learning techniques.

 \paragraph{Acknowledgments}

This work was partially funded 
by 
the German Science Foundation (DFG) under grants SCHA 550/8-2, SCHA 550/10-1 AOBJ: 593494.
%
%


\bibliographystyle{acmtrans}


\end{document}